\def\preprint{1}		%  preprint
\def\comment#1{}
\DeclareRobustCommand{\ion}[2]{%
\relax\ifmmode
\ifx\testbx\f@series
{\mathbf{#1\,\mathsc{#2}}}\else
{\mathrm{#1\,\mathsc{#2}}}\fi
\else\textup{#1\,{\mdseries\textsc{#2}}}%
\fi}
\if\preprint1
	\documentclass{mn2e}
	\usepackage{graphicx}
\else
	\documentstyle[astrop-bib,referee,times]{mn}
	\newcommand{\includegraphics}[1]{}
\fi

\def\oversim#1#2{\lower0.5pt\vbox{\baselineskip0pt \lineskip-0.5pt
     \ialign{$\mathsurround0pt #1\hfil##\hfil$\crcr#2\crcr\sim\crcr}}}
\title[High-velocity regions in planetary nebulae]
{High-velocity regions in planetary nebulae\thanks{Based on observations obtained 
at the European Southern Observatory.}}
   \author[Gesicki \&\ Zijlstra]{Krzysztof\,Gesicki$^1$\thanks{email: Krzysztof.Gesicki@astri.uni.torun.pl} 
		and
      Albert \,A.\,Zijlstra$^2$\thanks{email: a.zijlstra@umist.ac.uk}\\ \\
   $^1$ Centrum Astronomii UMK, ul.Gagarina 11, PL-87-100 Torun,
	     Poland. \\
   $^2$  UMIST, Department of Physics, P.O. Box 88,
           Manchester M60 1QD, UK.\\
 }

\pubyear{2002}

\begin{document}

\maketitle

\begin{abstract} 

The internal velocity fields of planetary nebulae are studied with a
resolution of 5\,km\,s$^{-1}$. We analyze deep echelle spectra from
three nebulae in the Bulge, the Sagittarius Dwarf and the SMC. No
effects of metallicity is seen, except possibly a slower onset of the
fast wind from the central star.  Robust evidence is found for the
existence of a high-velocity shock at the inner edges of the nebulae.
Such a shock is predicted in hydrodynamical models but had not
previously been observed. The shock gas is accelerated by the fast
wind from the central star. A similar shock at the outer edges traces
the expansion of the ionized shell into the ambient AGB wind.
Evidence for localized regions of high velocity is also found from
lines of intermediate excitation, for two of nebulae. We explore
several possible interpretations: (1) an embedded shock at
intermediate radii, as predicted by hydrodynamic models at the
position of the outer edge of the swept-up inner shell; (2) deviations
form spherical symmetry, where in some directions the
intermediate-excitation lines extend into the region of the outer
shock; (3) An intermediate swept-up shell, as seen in some Galactic
planetary nebulae.  The remaining nebula, with a [WC] star, shows
strong turbulence. This may trace a superposition of many embedded
shock-lets.  We suggest a relation to the time-variable [WC] wind,
giving a planetary nebula subjected to a multitude of sound waves.

\end{abstract}


\section{Introduction}

The evolution of planetary nebulae (PNe) is driven by several distinct
processes. The original density and velocity distribution are
determined by the AGB mass loss. Subsequently, this is modified by the
photo-ionization, and the interaction with the hot, fast wind from the
central star. Together, these give rise to the intricate structures
seen in many PNe.

The first models are appearing which use all of these processes to
calculate the instantaneous density and velocity structure
(e.g. Corradi et al. 2000a). The density structures can be obtained
from high-resolution imaging, but determining the dynamical structures
requires high-resolution spectra. We have developed a technique to
deconvolve the velocity field from the emission line profiles. Using
[\ion{O}{iii}] and [\ion{N}{ii}], this has shown that the velocity
tends to increase sharply towards the outer edge of the nebula, in
agreement with hydrodynamical models.  For PNe with [WC] central
stars, the lines show evidence for turbulent velocity fields (Acker et
al. 2002).

If a larger set of lines is available, the velocity field can be
obtained with much finer radial resolution, especially if narrow-band
images are also available. Such data is needed for comparison with the
details of the hydrodynamic models, and in this way can test the
evolution of PN nuclei and their mass loss history. Here we apply the
'Torun models' to deep echelle spectra of three PNe, located in the
Galactic Bulge, the Sagittarius Dwarf Spheroidal galaxy and the Small
Magellanic Cloud (SMC).  The new data allow us to deduce the velocity
fields with unprecedented detail. There are two main reasons for this
improvement: lines of many ionization stages are available, and the
spectral resolution of 60\,000 is well fitted to the thermal
broadening.

In the next sections, after a summary of the Torun models, we present
the results for three objects. Evidence for several shock-like features
is found. Although the objects were selected based on their different
metallicity, they are remarkably similar.  The core masses are
calculated using the dynamical age: they are all close to
0.62\,M$_\odot$.  We discuss the various shocks and interpret these in
terms of structures seen in the hydrodynamical models. One object
shows evidence for strong turbulence: we discuss the relation between
this and the shocks seen in the other nebulae.

\section{Observations}

We selected three PNe from different stellar populations: the Bulge,
the Sagittarius Dwarf Spheroidal galaxy and the SMC. These have the
advantage that the distances are approximately known, which benefits
the photo-ionization modeling. Diameters are known from published
monochromatic images for all three nebulae. One of our interests was
the effect of metallicity: values for the metallicity have been
published for these PNe, ranging from slightly subsolar (Bulge PN) to
$-0.6$ dex (SMC PN). All three central stars are hot with very similar
luminosity.

The observations were carried out at the ESO 3.5m NTT telescope, in
August 1999.  We used the echelle mode in the red arm, using grating
\#14 and cross disperser \#3. The slit-width was 1$\arcsec$\ and
slit length 3$\arcsec$. The spectra cover the wavelength range
4300--7000\AA\ at a resolution of $R= 6 \times 10^4$, corresponding to
5\,km\,s$^{-1}$: this is about the thermal line width for oxygen.  The
spectra were summed along the slit, flat-fielded and corrected for the
response curves, using the MIDAS echelle reduction package.  Each
source was observed for $3 \times 1800\,$sec. The rms noise of the
final 1-d spectra is approximately $10^{-4}$ of the H$\beta$ flux.

For the analysis we selected several (8--9) strong and well exposed
lines which cover a broad range of excitation potentials.  This allows
us to probe the whole nebular depth from the highly ionized inner
regions (\ion{He}{ii} 4686\AA) to the neutral outer layers
([\ion{O}{i}] 6302\AA).  The three pairs \ion{O}{iii} and
\ion{Ar}{iv}, \ion{S}{ii} and \ion{N}{ii}, and \ion{Cl}{iii} and
\ion{S}{iii}, each come from very similar region, leaving effectively
five independent probes.  The SMC PN was unresolved but the other two PNe
show line splitting in the low-excitation lines, which indicates they
are partially resolved. 

\section{The model analysis}

\subsection{The computer codes}

We applied the 'Torun models' (Gesicki et al. 1996) to deduce the
density and velocity radial distribution of our PNe. A brief
description of the procedure is repeated here. We emphasize that
the models assume spherical symmetry (as do most photo-ionization models)
and this is likely to be the limiting approximation.

The central star is approximated by a black-body atmosphere defined
by an effective temperature and a luminosity.  The nebula is
approximated as a spherical shell defined by an inner and an outer
radius, a total mass and a radial density distribution.  A
photo-ionization model is calculated to fit the (published) line
ratios, H$\beta$ flux, and electron density and temperature.  We adopt
published values for the metallicity.  The model is used to calculate
predicted surface brightness profiles, which are compared to published
monochromatic images. The model is iterated until an acceptable
agreement is found: the main 'free' parameters in this iterating
procedure are the radial density distribution and the central star
parameters. For every line under consideration, a radial emissivity
distribution is calculated.

Subsequently a radial velocity profile is assumed, with the velocity
varying smoothly with radius.  At every radial position, the line
profile is calculated from the local velocity (and turbulence), the
local electron temperature and the local emissivity.  For every point
on the image of the nebula, we calculate a total line profile by
integrating along the line of sight. The observed line profile is
obtained by masking the model nebula using the slit parameters, and
summing over this slit. The seeing is taken (approximately) into
account as an additional broadening of the slit.  Comparing the
predicted profile with the observations now allows one to correct the
assumed velocity field. In addition to the velocity profile, a
turbulent component can also be included. New to the Torun models is
the possibility to allow for a radially varying degree of turbulence.

In the Torun models, the radial velocity is specified at an arbitrary
number of positions. These are fitted using smoothed cubic splines.
In previous papers, relatively simple velocity profiles were
applied. Here we investigate the velocity fields in much more detail
than has been possible before.

\subsection{The applicability of models}

Because spherical symmetry is built into our (photo-ionization) models
we restrict the analysis only to those objects which are not too far
away from appearing circular. Although many PNe show very complicated
structures, both in images and spectra, near-spherical PNe still
constitute a significant subset of all PNe with known
shapes. Depending on the definition their fraction is estimated as
between 10\% and 25\% (Soker 2002). The fraction of objects suitable
for our analysis can be much broader if we include moderately
elliptical PNe.

An example of a prolate ellipsoidal PN is NGC\,7027, with an axial
ratio of 2:1. Long-slit echelle observations show that the same
velocity law applies to both nebular axes (Walsh et al. 1997a).
Hydrodynamical calculations of aspherical PNe (Mellema 1995) find
uniform expansion in all directions. These observational and
theoretical findings suggest that our spherical velocity modelling can
be applied to mildly elliptical nebulae, which constitute of the order
of 50\%\ of known objects. Bipolar objects, which can have
high-velocity polar flows, should be avoided especially when
integrated spectra are used.  The presence of bipolar flows is often
evident from integrated spectra. In earlier applications of the Torun
models objects with marked bipolarity were always excluded.

The velocity accuracy of our method is restricted by the spectral
resolution of the observations, limited by the spectrograph and the
thermal line widths. In the present work we cannot resolve details
smaller than about 5\,km\,s$^{-1}$. The spatial resolution of our
modelling is limited by the apparent size of the PNe which is comparable
to both the size of the slit and the seeing.  Because of these
observational limitations the analysis must be restricted to large
scale structures and flows, and small inhomogeneities cannot be
addressed. This also argues against attempting bipolar nebulae.

In the next sections, describing each single PN, we discuss the
available images. There is some evidence for deviations from
sphericity: Wray\,16-423 may show elongation, SMP\,5 is circular but
could be a flattened shell seen face on (suggested by the fact that
the central brightness is not well fitted), and M 2-31 appears
circular. We find no evidence of marked bipolarity in images, and the
regular line profiles show no indications of (bi-)polar flows.

Using a radiation-hydrodynamic code, Mellema (1997) suggests that
bipolar nebulae much easier evolve around higher mass central stars
(0.836 M$_{\odot}$) while lower mass stars (0.605 M$_{\odot}$) tend to
produce elliptical PNe. This is because the less massive star evolves
slower, giving the density distribution in the surrounding
shell time to smooth out. Anticipating our results presented later, 
the fact that our central stars have masses close to the lower value
favours an elliptical shape over a bipolar one.

\subsection{The modelling procedure}

The velocity field is obtained by trial and error; we cannot be certain
that the solutions are unique. Nevertheless we always start with the
simplest possible field, i.e. constant velocity.  While comparing the
computed profiles with the observed ones, taking into account the
earlier calculated ion stratifications, we change the velocities at
different radial distances to obtain the required change in
profiles. We always search for the simplest solution which improves as
much as possible all reliable residual in line shapes. We illustrate
this procedure in detail when discussing the first object; for the
other two PNe we skip this step and show only the final results.

Tables\,\ref{taba} and \ref{tabb} summarize the input parameters of
the three modelled PNe. They also list some models results and compare
them with literature data. For the comparison with the observed
$T_{\rm e}$ and $N_{\rm e}$, we list the model values at the radii at
which the emissivity of the relevant ion peaks. $V_{\rm av}$ denotes
the mean value of the velocity field weighted by the mass distribution
(Gesicki et al. 1998). The metallicity listed as [O/H] in fact
presents the average of [Ne/H] and [O/H]: neon follows oxygen very
closely in the ISM (Henry 1989) with identical production sites, but
oxygen in PNe can be affected by 3rd dredge of primary oxygen,
especially at low metallicity (P\'equignot et al. 2000).

The individual objects are discussed extensively in the following
sections, where literature references are given. Each section begins
with a description of the nebula, followed by subsections on the
density distribution and the velocity field.

%%%%%%%%%%%%%%%%%%%%%%%%%%%%%%%%%%%%%%%%%%%%%%%%%%%%%%%%%%%%%%%%%%%%%%%
\begin{table*}
\caption[]{Parameters of the nebular models compared with observed
values}
\begin{flushleft}
\begin{tabular}{ l l l l l }
\hline \noalign{\smallskip} object & M\,2-31 & SMP\,5 & Wray\,16-423
\\ 
\noalign{\smallskip} & PN\,G\,6.0-3.6 & & PN\,G\,6.8-19.8\\
\noalign{\smallskip} 
\hline 
\noalign{\smallskip} distance [kpc] &
\multicolumn{3}{c}{ } \\ 
\hskip1cm from literature & 4.75 & 60.3 & 25
\\ 
\hskip1cm assumed in model & 8.0 & 60.3 & 25 \\
\noalign{\smallskip} & \multicolumn{3}{c}{ } \\ 
\noalign{\smallskip}
NEBULA & \multicolumn{3}{c}{ } \\ 
Angular diameter [ arc sec ] &
\multicolumn{3}{c}{ } \\ 
\hskip1cm from literature: & opt. 2.1, radio
3.8 & opt. 0.5 & opt. 1.2, radio 0.7$\times$0.3 \\ 
\hskip1cm from
model H$\beta$ image & 2 core, 3.6 halo & 0.5 core, 1 halo & 1 core,
1.5 halo \\ 
model outer radius & 2.2\,10$^{17}$cm (0.07\,pc) &
4.7\,10$^{17}$cm (0.15\,pc) & 2.8\,10$^{17}$cm (0.09\,pc) \\ 
model
inner radius & $0.16\,10^{17}$\,cm & $0.47\,10^{17}$\,cm &
$0.28\,10^{17}$\,cm \\ 
extinction c(H$\beta$) & 1.34 & 0.097 & 0.21 \\
\noalign{\smallskip} log F(H$\beta$) & \multicolumn{3}{c}{ } \\
\hskip1cm observed, dereddened & -10.57 & -12.65 & -11.89 \\ 
\hskip1cm
calculated from model & -10.6 & -12.56 & -11.73 \\
\noalign{\smallskip} [\ion{O}{iii}] electron temperature [K] &
\multicolumn{3}{c}{ } \\ 
\hskip1cm from literature & 10100, 9890 &
13629 & 12400 \\ 
\hskip1cm from our models & 9740 & 13100 & 12380 \\
\noalign{\smallskip} [\ion{N}{ii}] electron temperature [K] &
\multicolumn{3}{c}{ } \\ 
\hskip1cm from literature & 11400, 9775 & --
& \\ 
\hskip1cm from our models & 10400 & 11600 & 12350 \\
\noalign{\smallskip} [\ion{S}{ii}] electron density [cm$^{-3}$] &
\multicolumn{3}{c}{ } \\ 
\hskip1cm from literature & 4030, 5050 & -- &
6000 \\ 
\hskip1cm from our models & 4490 & 790 & 1290 \\
\noalign{\smallskip} $V_{\rm exp}$ [km\,s$^{-1}$] observed &
\multicolumn{3}{c}{ } \\ 
\hskip1cm [\ion{O}{iii}] peak separation & 9
& -- & -- \\ 
\hskip1cm [\ion{O}{iii}] HWHM & 25 & 19 & 23 \\ 
\hskip1cm
[\ion{N}{ii}] peak separation & 23 & -- & 21 \\ 
\hskip1cm
[\ion{N}{ii}] HWHM & 35 & 23 & -- \\ 
\noalign{\smallskip} $V_{\rm av}$
[km/s] calculated from model & 30 & 30 & 33 \\ 
additional turbulent
broadening [km/s] & 10--20 & 0 & 0 \\ 
\noalign{\smallskip} ionized
mass [M$_{\odot}$] calculated from model & 0.26 & 0.61 & 0.33 \\
\noalign{\smallskip} metallicity [O/H] & $-0.2:$ & $-0.6$ & $-0.55$ \\
\noalign{\smallskip} & \multicolumn{3}{c}{ } \\ 
\noalign{\smallskip}
STAR & \multicolumn{3}{c}{ } \\ 
\noalign{\smallskip} spectrum
description & [WC 4-6], wels & -- & [WC 4--6], wels \\ 
log($T_{\rm
eff}$) from literature & 4.8, 4.9 & 5.14 & 4.95 or 5.1 \\ 
log($T_{\rm
eff}$) adopted & 4.86 & 5.14 & 5.0 \\ 
log($L/L_{\odot}$) from
literature & 3.6, 3.5 & 3.77 & 3.66 or 3.7 \\ 
log($L/L_{\odot}$)
adopted & 3.7 & 3.61 & 3.7 \\ 
\noalign{\smallskip} estimated dynamical
core mass [M$_\odot$] & 0.62 & 0.61 & 0.61 \\ 
\noalign{\smallskip}
\hline
\end{tabular}
\end{flushleft}
\label{taba}
\end{table*}

%%%%%%%%%%%%%%%%%%%%%%%%%%%%%%%%%%%%%%%%%%%%%%%%%%%%%%%%%%%%%%%%%%%%%%%

\begin{table*}
\caption[]{Comparison of observed (dereddened) and calculated line
ratios, relative to I(H$\beta$) = 100}
\begin{flushleft}
\begin{tabular}{ l l l l l l l l l l l }
\hline \noalign{\smallskip} & \multicolumn{3}{c}{M\,2-31} & &
\multicolumn{3}{c}{SMP\,5} & & \multicolumn{2}{c}{Wray\,16-423} \\
\noalign{\smallskip} & \multicolumn{3}{c}{PN\,G\,6.0-3.6} & &
\multicolumn{3}{c}{ } & & \multicolumn{2}{c}{PN\,G\,6.8-19.8} \\
\noalign{\smallskip} \cline{2-4}\cline{6-8}\cline{10-11}
\noalign{\smallskip} & \multicolumn{2}{c}{observed} & model & &
\multicolumn{2}{c}{observed} & model & & \multicolumn{1}{c}{observed}
& model \\ line & (1) & (2) & & & (2) & (3) & & & (4) & \\
\noalign{\smallskip} \hline \noalign{\smallskip}

  [\ion{O}{ii}] 3726 & -- & 66 & 53 & & 80 & -- & 55 & & 19 & 45 \\

  [\ion{Ne}{iii}] 3868 & -- & 103 & 132 & & 87 & -- & 69 & & 74 & 83
                       \\

   \ion{He}{ii} 4686 & -- & 0 & 3 & & 41 & 41 & 35 & & 11 & 13 \\

  [\ion{O}{iii}] 5007 & 1116 & 973 & 910 & & 962 & -- & 916 & & 1085 &
                         997 \\

  [\ion{O}{i}] 6300 & 6 & 4 & 4 & & 9 & -- & 4 & & 2 & 2 \\

  [\ion{S}{iii}] 6311 & 3 & -- & 2 & & -- & -- & 1 & & 2 & 2 \\

  [\ion{N}{ii}] 6584 & 60 & 77 & 93 & & 25 & -- & 22 & & 18 & 37 \\

  [\ion{S}{ii}] 6717+6731 & 13 & 11 & 13 & & 9 & 8 & 8 & & 5 & 6 \\

  [\ion{Ar}{iii}] 7136 & 17 & -- & 12 & & -- & -- & 6 & & 10 & 5 \\
\noalign{\smallskip} \hline \noalign{\smallskip}
\multicolumn{3}{l}{(1) Acker et al. (1991) } & \multicolumn{8}{l}{(2)
Stasinska et al. (1998) } \\ \multicolumn{3}{l}{(3) Liu et al. (1995)
} & \multicolumn{8}{l}{(4) Walsh et al. (1997b) } \\
\end{tabular}
\end{flushleft}
\label{tabb}
\end{table*}

%%%%%%%%%%%%%%%%%%%%%%%%%%%%%%%%%%%%%%%%%%%%%%%%%%%%%%%%%%%%%%%%%%%%%%%%%%

\section{Wray\,16-423 (PN\,G\,006.8$-$19.8)}

   \begin{figure*}
   \resizebox{\hsize}{!}{\includegraphics{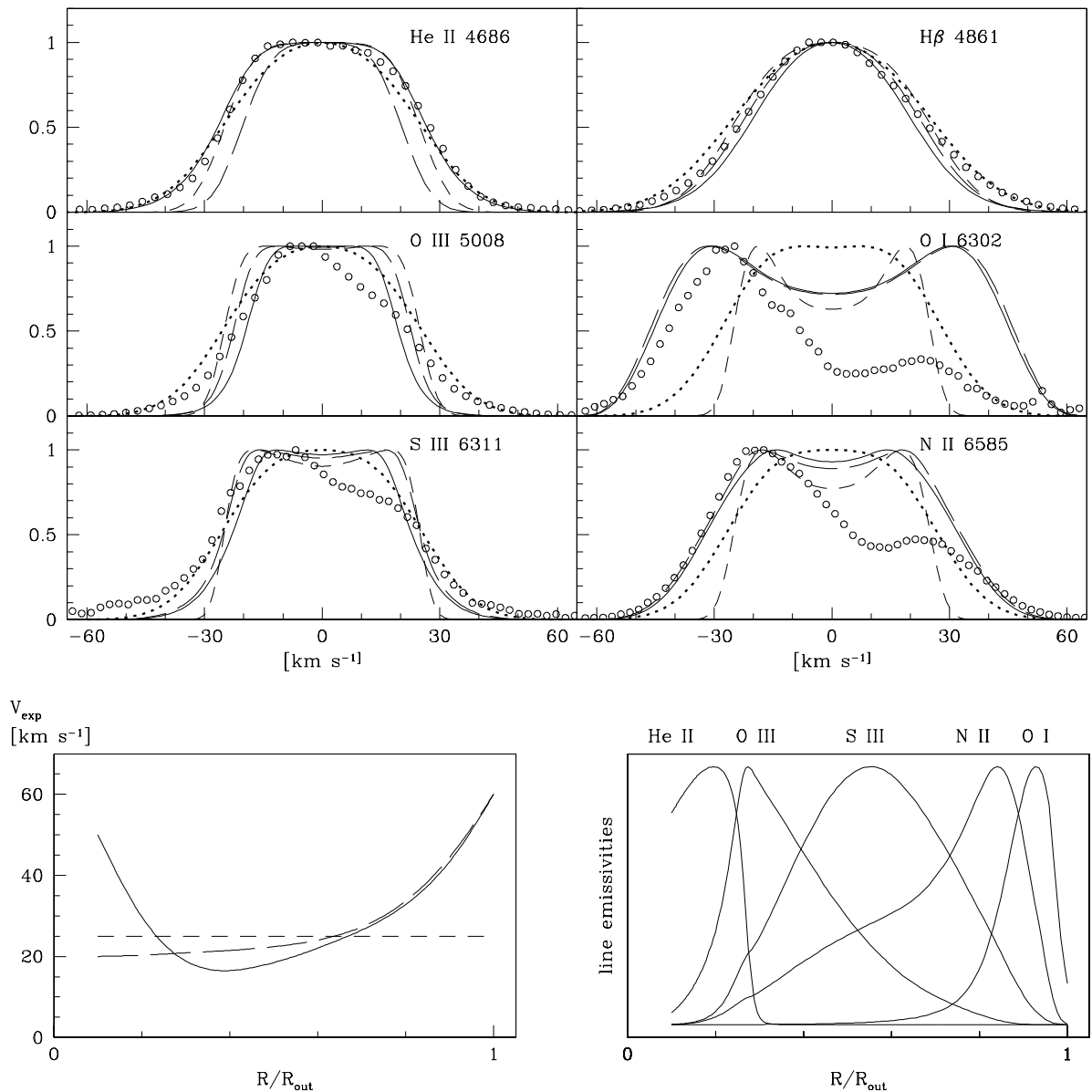}}
   \caption{Examples of rejected models: the observed and modelled
   lines and model parameters for Wray\,16-423, using over-simplified
   velocity fields.  The upper panels present six representative
   emission lines; the points correspond to the observations, and the
   lines to the calculated models.  The intensities are normalized to
   unity. The velocity scale is given in the lowest boxes.  In the
   lower left panel, the different velocity fields are marked with
   different line types, corresponding to the emission lines shown
   above. The right panel shows the radial emissivity distribution in
   selected lines. More explanation is given in the text.}

\label{bad_fit} 
\end{figure*}

   \begin{figure*} \resizebox{\hsize}{!}{\includegraphics{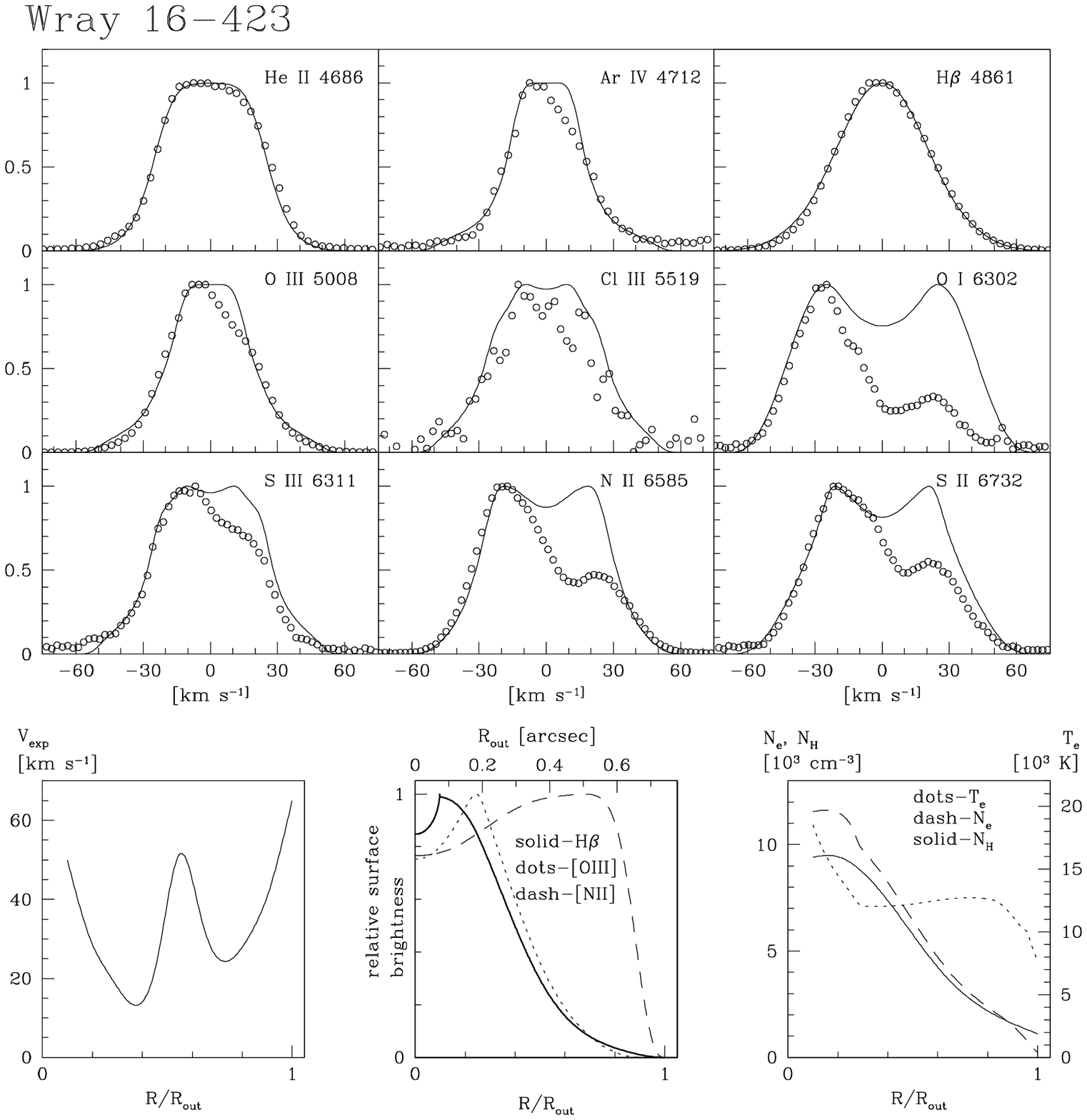}}
  \caption{ Observed and modelled lines and model parameters for
  Wray\,16-423.  The upper panels present the nine emission lines; the
  points correspond to the observations, and the solid curves show the
  calculated models.  The intensities are normalized to unity. The
  velocity scale is given in the lowest boxes.  The lower panels show
  the model structure as function of the relative radius.  In the left
  panel, the velocity field is shown.  The central panel shows the
  model surface brightness profiles.  The right panel shows the
  assumed density distribution together with the model electron
  temperature and density curves.  } 
\label{fig_w} 
\end{figure*}

\subsection{The nebula}

The planetary nebula Wray\,16-423 is located at a distance of 25\,kpc
in the Sagittarius Dwarf Galaxy (Zijlstra \&\ Walsh 1996). The PN has
a well-determined progenitor mass (1.3 M$_\odot$) and metallicity
([Fe/H]=$-0.55$). The central star shows a weak [WC] (wels)
spectrum. The other known PN in this galaxy, He\,2-436, shows strong
[WC] lines (Walsh et al. 1997b). Originally, Wray\,16-423 was
classified as [WC8], but Dudziak et al. (2000) show that this was
based on a wrong \ion{C}{iii} identification: the type is considerably
earlier, in better agreement with the high effective temperature.

A photoionization model study of Wray\,16-423 has been published by
Dudziak et al.  (2000), where a detailed chemical composition is
derived.  Their final model which fits all observed line ratios is a
composite of optically thin and thick components.  We adopt their
chemical composition; some of their data are given in
Table\,\ref{taba}.

For our work we selected 9 spectral lines (presented by open circles
in Fig.\,\ref{fig_w}). Some of these, especially [\ion{O}{i}],
indicate the presence of an asymmetry in this PN.  \ion{O}{iii} and
radio images (Dudziak et al. 2000) indicate a partially resolved,
elliptical nebula, with a major axis around 1\arcsec\ and a (more
uncertain) minor axis of 0.7\arcsec.

\subsection{Size and density distribution}

The observed angular radius of 0.5\arcsec (Walsh et al. 1997b), at the
distance of 25\,kpc, corresponds to 0.06\,pc.  The high excitation
lines (\ion{He}{ii}, [\ion{Ar}{iv}], [\ion{O}{iii}]) do not show line
splitting, confirming that the (inner) emitting region is smaller than
the slit width.  However, lower-excitation lines do show some line
splitting, and must form in a more extended region.  We adopt an outer
radius of 0.09\,pc (0.75\arcsec), and a density structure for which
the calculated surface brightness in H$\beta$ decreases to 10\% of
maximum at 0.5\arcsec.  Our photoionization code gives line ratios
close ($\sim 50\%$) to the values given in Dudziak et
al. (2000). However, there is a noticeable difference between the
electron density estimated from observed [\ion{S}{ii}] line ratios
(Walsh et al. 1997b) and that in the outer layers of our model.
Dudziak et al. (2000) model Wray 16-423 with two sectors: for their
lower-density sector they find $N_{\rm e} = 3600\rm
\, cm^{-3}$ which is in between our and Walsh et al. values.  A
high-resolution image of a low-excitation line could resolve this
issue, but no such image is presently available.
The model is ionization bounded.

\subsection{Velocity fields}

We begin the discussion of the first nebula with an overview of simple
velocity fields, to illustrate the procedure to obtain an acceptable
model. The results of the initial models are shown in
Fig.\,\ref{bad_fit}. We show only six lines in the plot, nevertheless
the chosen lines represent the whole nebula. The
constant-velocity-field model is presented in two versions: without
and with additional turbulence. The short-dash line presents the
non-turbulent solution: the value of 25\,km\,s$^{-1}$ is on average
the best value. To show how additional turbulent motions change the
line shapes we compute the same model with a very high turbulence of
15\,km\,s$^{-1}$. The dotted line presents this exaggerated case; it
shows that the line wings are broadened and simultaneously the maxima
are smoothed to a Gaussian shape. Both cases fail for almost all
lines.

Before modifying the velocity we look at the lower right panel of
Fig.\,\ref{bad_fit} showing the radial distributions of emissivities.
Again not all nine lines are presented because several distributions
almost overlap and H$\beta$ follows closely the density shown in the
next panel. Fig.\,\ref{bad_fit} indicates the radii of formation of
considered lines, i.e. the radii where the line profiles are most
sensitive to changes in velocity.

The largest discrepancies exist for low excitation lines, indicating a
required increase of the velocity in the outer region. The long-dash
line shows the (non turbulent) monotonic velocity field which usually
was assumed (Gesicki et al. 1998) and which is predicted from
hydrodynamic models (Marten \&\ Szczerba 1997, Perinotto et al. 1998).
The new fit shows clearly that this is the appropriate way to
reproduce the low excitation outermost lines.  The acceleration in the
outermost layers is connected with the ionization front,
and is a known, common feature of PNe (Acker at al. 2002).

As a new result, the monotonic velocity field also fails at the
highest excitations in Wray\,16-423. The \ion{He}{ii} line is too
broad: a high velocity (50\,km\,s$^{-1}$), high density region is
required near the inner nebular radius. Further out the velocity must
decrease significantly (to about 15\,km\,s$^{-1}$) to reproduce the
narrow cores of the [\ion{O}{iii}] and [\ion{Ar}{iv}] lines.  The
outer regions accelerate to 60\,km\,s$^{-1}$, and must show a density
decreasing with radius, otherwise the high velocity wings become too
pronounced. A constant density halo can be ruled out.

This velocity field is shown in Fig.\,\ref{bad_fit} as the solid
line. The general shape of all 9 lines is reproduced (see the Figure),
but it is clearly visible that a number of lines with intermediate
excitation ([\ion{S}{iii}], [\ion{Cl}{iii}], [\ion{O}{iii}],
[\ion{Ar}{iv}]) have broader wings than the fit predicts. This shows
that a component expanding at about 40\,km\,s$^{-1}$, at intermediate
radii, is missing.  Adding this additional high velocity component
improves the fit significantly.  The resulting velocity profile, and
the resulting emission line profiles, are shown in Fig.\,\ref{fig_w}.
The assumed velocity components have an almost negligible effect on the
H$\beta$ line because (being least massive) hydrogen is strongly
thermally broadened.

The 'best model' gives an interesting velocity profile.  The velocity
spike at $R \approx 0.5R_{\rm out}$ gives the impression of an
embedded shock.  If the density is also enhanced in this region, the
high velocity region would become even thinner.  Alternative
possibilities are explored in Section 10.

%%%%%%%%%%%%%%%%%%%%%%%%%%%%%%%%%%%%%%%%%%%%%%%%%%%%%%%%%%%%%%%%%%%%%%%%%%

\section{SMP\,5}

\subsection{The nebula}

This planetary nebula is located in the Small Magellanic Cloud.  A
photoionization model is published by Liu et al.  (1995), based on
optical and UV spectrophotometry and on Hubble Space Telescope images.
The data given in Table\,\ref{taba} and Table\,\ref{tabb} are taken
from this paper.  Their model reproduces many observed parameters;
however, it fails to fully reproduce the [\ion{O}{iii}] image: the
calculated [\ion{O}{iii}] surface brightness decreases faster outwards
than observed.  Another photoionization model is published by
Stasinska et al. (1998); their data supplement our tables.  The
chemical composition is adopted from their paper.

\begin{figure*} \resizebox{\hsize}{!}{\includegraphics{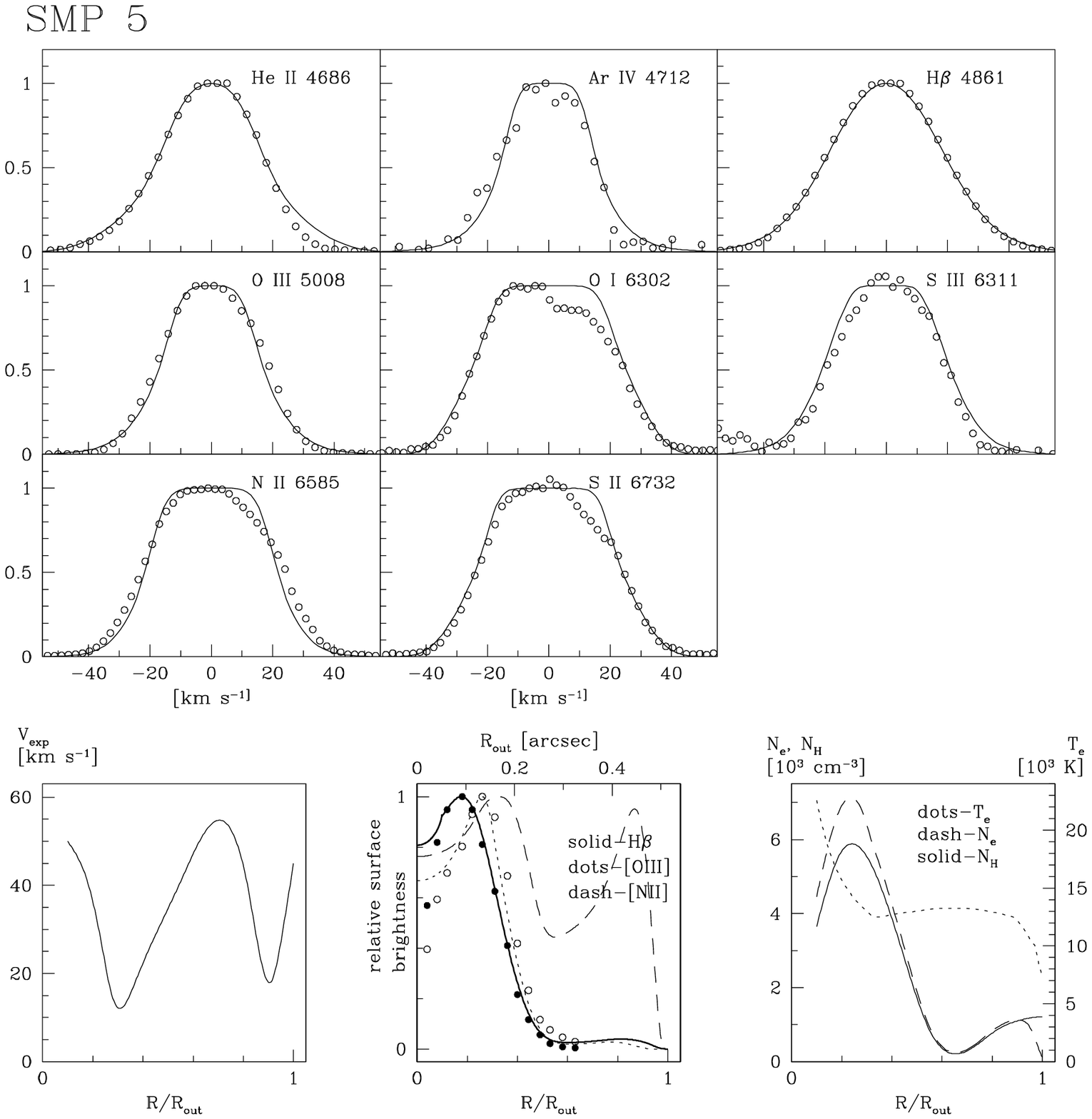}}
   \caption{ Observed and modelled lines together with model
   parameters of SMP\,5.  The arrangements are the same as in
   Fig.\,\ref{fig_w}, but only a single velocity solution is shown,
   and in the relative surface brightness box the observed points are
   overplotted.  }  \label{fig_s} \end{figure*}

In our echelle spectrum, 8 lines are suitable for model analysis.  The
observed line profiles are presented by the open circles in
Fig.\,\ref{fig_s}.

\subsection{Size and density distribution}

The H$\beta$ and [\ion{O}{iii}] 5007\AA\ images (Liu et al.  1995)
show a nearly circular nebula. The azimuthally averaged surface
brightness distributions of Liu et al.  are shown in Fig.\,\ref{fig_s}
(lowest central box), where the filled circles correspond to H$\beta$
and the open circles to [\ion{O}{iii}].  We first tried to reproduce
these surface brightness profiles by iterating (trial and error) the
density structure. The result is shown in the same figure, together
with the predicted \ion{N}{ii} distribution.

The H$\beta$ image shows an outer radius of 0.25\arcsec. The
[\ion{O}{iii}] image is somewhat larger, and peaks further out than
H$\beta$ by about 0.04\arcsec. The reason is that the star is
sufficiently hot that the innermost region is dominated by \ion{O}{iv}
(e.g.  Leene \&\ Pottasch 1987), up to the density maximum at
0.1\arcsec\ radius.  The observed [\ion{O}{iii}] emission remains
brighter than H$\beta$ in the outer regions.  This can be expected
in density-bounded nebulae, where the \ion{O}{iii} zone extends to the
outer radius.  But in such density-bounded models, the [\ion{O}{i}]
6300\AA\ line intensity is about $10^{-4}$ of H$\beta$, much fainter
than observed.  SMP5 appears to be ionization-bounded.

To fit both the extended [\ion{O}{iii}] distribution and the strength
of the [\ion{O}{i}] line, we appended a low-density outer region
extending out to a distance of 0.5\arcsec. In this model, the
[\ion{O}{iii}] zone terminates at a radius of 0.4\arcsec.  Placing the
ionization front still farther out could reproduce the tail of the
[\ion{O}{iii}] distribution even better, but its radius cannot be much
larger than 0.5\arcsec: otherwise the [\ion{O}{i}] 6300\AA\ line would
be expected to show line splitting, which is not observed.

Increasing the stellar $T_{\rm eff}$ would shift the [\ion{O}{iii}]
image still farther out but it is not an overwhelming improvement: we
adopted the temperature of Liu et al.  (1995).  We obtain a better fit
to the brightness distribution than Liu et al.: they use a slightly
higher density than ours, resulting in an smaller ionization radius.
In our model, the outer region should be visible in the 6585\AA\
[\ion{N}{ii}] line.  An image at this wavelengths could decide between
the two models. Without such an image, the halo is in essence a free
parameter in our modelling.  We note that our model fails to fit the
image details in the very central region.

A consequence of the assumed outer region is the high nebular mass
of 0.61\,M$_{\odot}$, much higher than 0.194\,M$_{\odot}$ obtained by
Liu et al.  (1995). Nevertheless the H$\beta$ flux and the line ratios
are well reproduced.

\subsection{Velocity field}

The different emission lines show different line widths.  Assuming the
halo has constant density, or slightly decreasing like in
Wray\,16-423, causes the maxima of the emissivity distribution of all
lines to be contained in the dense central region. But the different
line widths indicate the line emission regions are well separated.
Introducing a small density increase outwards shifts the low
excitation lines outwards, as required.  The \ion{N}{ii}, \ion{S}{ii},
\ion{S}{iii} ions now exhibit two emissivity maxima: one connected with
the density maximum, the other with the ionization stratification in
the outer region.

The \ion{He}{ii} line is broader than [\ion{O}{iii}] and
[\ion{Ar}{iv}] which implies an increase of the velocity towards the
central star. The width of the [\ion{O}{i}] line indicates
acceleration at the outer nebular radius. A flat-top line profile
results from a spatially unresolved layer expanding with constant
velocity (Gesicki \& Zijlstra 2000).  The narrow flat tops of the
[\ion{O}{iii}] and [\ion{Ar}{iv}] lines indicate the presence of such
layers in the inner dense layers, with a low expansion velocity of
about 15\,km\,s$^{-1}$.  Similar flat tops in the [\ion{O}{i}],
[\ion{N}{ii}], [\ion{S}{ii}] lines indicate another velocity minimum
at about 20\,km\,s$^{-1}$ in the outer layers.  The line wings, which extend
over 50\,km\,s$^{-1}$, require a velocity maximum in between.  Any
velocity field simpler than shown cannot reproduce all selected lines
simultaneously.

%%%%%%%%%%%%%%%%%%%%%%%%%%%%%%%%%%%%%%%%%%%%%%%%%%%%%%%%%%%%%%%%%%%%%%%%%%

\section{M\,2-31, (PN\,G\,006.0-03.6)}

\subsection{The nebula}

   \begin{figure*} \resizebox{\hsize}{!}{\includegraphics{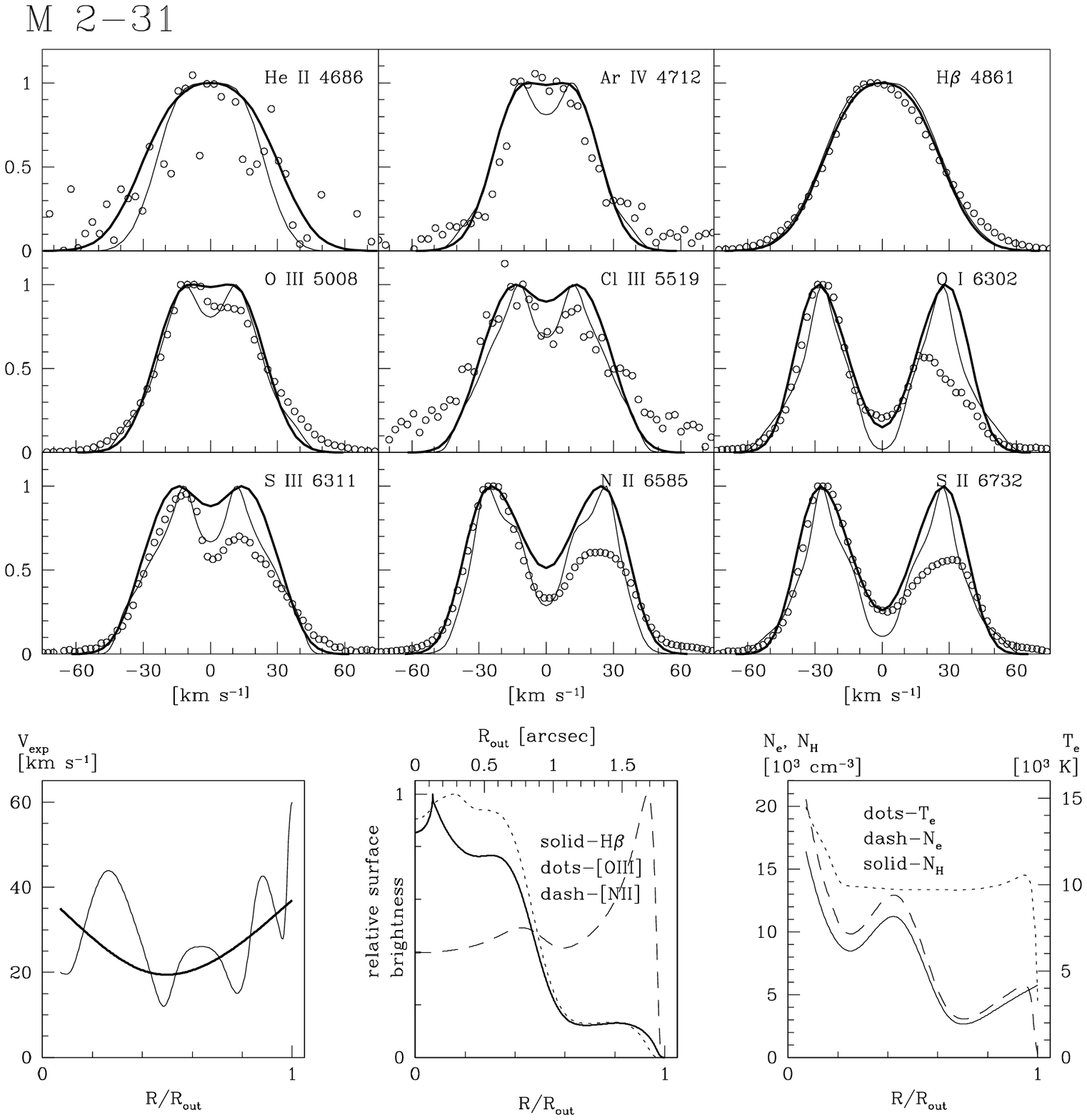}}
   \caption{ Observed and modelled lines together with model
   parameters of M\,2-31. The arrangements are the same as in
   Fig.\,\ref{fig_w}.  } \label{fig_m} \end{figure*}

This PN is likely located in the Galactic Bulge. Van de Steene \&
Zijlstra (1994) estimated the distance as 4.75\,kpc.  Dopita et
al. (1990) adopted 7.8\,kpc assuming that M\,2-31 is a Galactic Bulge
nebula. We obtain a better fit using this larger distance (see below).

The central star is classified as [WC 4--6] by Tylenda et al. (1993)
but as 'wels' by Parthasarathy et al. (1998).  PNe with [WC]-type
central star show a tendency towards strong turbulent motions: the
evidence for this comes from Gaussian line shapes which are of
comparable width for all species (Gesicki \&\ Acker 1996, Acker et
al. 2002)\footnote{A Gaussian shape for a single line can be fitted using
radially increasing velocities, but this does not produce the same
width for all lines.}. The first evidence for turbulence was based on
line profiles of [\ion{N}{ii}], H$\alpha$ and [\ion{O}{iii}] (Gesicki
\& Acker 1996).  Additional lines later yielded much stronger
constraints on the presence of turbulence (Neiner et al. 2000). We
cover similar lines to Neiner et al., but at 50\%\ higher resolution
and with higher S/N.

Photoionization studies of M\,2-31 have been published by Acker et
al. (1991) and Stasinska et al. (1998); their results are given in
Table\,\ref{taba} and Table\,\ref{tabb}. The chemical composition is
adopted from Stasinska et al. (1998). The abundances may be more
uncertain than in the previous two PNe: the [Ne/H] is about solar
while the [O/H]$ = -0.4\,\rm dex$ (see Stasinska et al. 1998): Ne and
O are expected to follow closely similar abundance abundance patterns
(Pequignot et al. 2000, Henry 1989) and the large difference would
need confirmation.

A very high-resolution [\ion{O}{iii}] 5007\AA\ image, and
spectrophotometry data are published by Dopita et al. (1990); they
also present a photoionization model analysis. A radio image is
published by Gathier et al. (1983).

For our analysis we selected 9 spectral lines, presented by open
circles in Fig.\,\ref{fig_m}. We include the \ion{He}{ii} line despite
its low signal to noise: it is the only available probe for the
innermost nebular layers.

\subsection{Size and density distribution}

We first attempted to obtain a photoionization model using the smaller
statistical distance (Van de Steene \&\ Zijlstra 1994): this was
rejected because to fit the line ratios and H$\beta$ flux, the model
nebula has a very small ionized mass and the central star a very low
luminosity.  This can not be {\it a priori} excluded, but at the
Galactic Bulge distance these parameters become more regular (see
Table\,\ref{taba}) and this was therefore considered more likely.

The assumed distance determines the nebular physical radius.  The
[\ion{O}{iii}] 5007\AA\ image of Dopita et al.  (1990) shows a
diameter of about 2.2\arcsec\ and is flat-topped, without a central
hole, and with steep edges.  The radio image of Gathier et al.  (1983)
shows the brightness increasing steadily towards the center; the
diameter is estimated as 2.1\arcsec.  Zijlstra et al. (1989) estimated
from the same image the diameter as 3.8\arcsec.

The dimensions of the spectrograph slit were $3\arcsec\times1\arcsec$.
In the observed spectrum the high excitation lines are unresolved
while the low excitation lines are clearly resolved.  This confirms
the presence of ionization stratification, where the central nebular
regions (traced by \ion{O}{iii} and \ion{Ar}{iv}) are confined within
the slit width, while the outer regions, traced by the lower
ionization stages, extend beyond the slit.  We adopted an outer radius
of 1.8\arcsec\ but searched for a model producing a much smaller
[\ion{O}{iii}] image.

The observed structure can be modelled with the density distribution
shown in Fig.\,\ref{fig_m}.  To reproduce the [\ion{O}{iii}] image
without a central hole it is necessary to assume a very small radius
of the empty inner cavity, with density higher towards the center.  To
obtain a flat-topped image out to $0.4\,R_{\rm out}$, we need to also
increase the density outside the [\ion{O}{iii}] dominated region.  We
implemented this as a high density bump near $0.4\,R_{\rm out}$.  Now
the [\ion{O}{iii}] image is almost flat-topped (it shows some
structure at intensities $0.9 <I/I_{\rm peak} <1.0 $), and decreases
to 10\% of maximum near $0.6\,R_{\rm out}$ which is comparable to the
observations.

For $R > 0.7\,R_{\rm out}$ the density in the model increases again.
This assumption, similar to the case of SMP\,5, was needed to shift
the low excitation lines to the outer nebular regions to reproduce the
splitted line profiles.  Also, without such a density increase the
calculated [\ion{O}{i}] 6300\AA\ line would be too weak, compared to
the observed line ratios.  These remote layers do not enlarge the
[\ion{O}{iii}] image.  The outer nebular radius must correspond to an
ionization front: in a density-bounded model the lines of ions
[\ion{O}{i}], [\ion{N}{ii}], [\ion{S}{ii}] become too weak.

\subsection{Velocity field}

The line emissivity distributions show complicated radial structure.
The high excitation lines show two closely-spaced emission peaks,
corresponding to the two inner density maxima.  The outermost low
excited lines emit mostly near the outer radius.  The 
intermediate-excitation ions \ion{Cl}{iii} and \ion{S}{iii}  exhibit two
well-separated emission maxima: they are not affected by the density
maximum at the inner edge.

No simple velocity field can explain the observed lines. Being
encouraged by the shocked velocity field of SMP\,5 we tried to find
something similar.  Without additional turbulence, the best fit for
all 8 lines can be achieved with the 'wavy' velocity curve presented
by the thin line in Fig.\,\ref{fig_m}.  The very outer velocity
increase is again naturally explained by the presence of the ionization
boundary. The innermost velocity is not probed even by the
\ion{He}{ii} ion. We assumed a velocity decrease in this region,
which could be associated with the strong increase of density
towards the center, and is also consistent with the not very broad
(but noisy) \ion{He}{ii} 4686\AA\ line.

A concern of this fit is that the number of extrema in the velocity
field is similar to the number of independent probes (see Section 2).
Fig.\,\ref{fig_m} shows some residuals are still present in the modelled
lines despite the elaborate velocity profile.  The observed emission
lines are clearly smooth in shape: higher-order components would be
needed in the velocity field to fully reproduce the spectral lines.
Such a situation occurs more naturally when the velocity field is
dominated by turbulence.  Therefore we searched for another best fit
model, assuming the simplest possible velocity field and including
turbulent broadening of lines.

We cannot find a satisfactory fit with the linear velocity field found
for many WR planetaries (Acker et al. 2002): the [\ion{Ar}{iv}] and
[\ion{O}{iii}] lines show clearly a smaller expansion velocity than
[\ion{O}{i}], [\ion{N}{ii}], [\ion{S}{ii}]. The best and the simplest
fit is obtained with a parabola-like velocity field, shown in
Fig.\,\ref{fig_m} by the thick line. The expansion velocity must be
near 20\,km\,s$^{-1}$ at the intermediate radii, while the
[\ion{O}{i}] line indicates a velocity increase outwards, and the
extended wings of the [\ion{O}{iii}] lines indicate an increase
inwards.  The best fit is obtained with a turbulence which is also
varying, decreasing from 20\,km\,s$^{-1}$ at the inner radius to
10\,km\,s$^{-1}$ at the outer edge.

Neither velocity field gives as good a fit as found for  SMP\,5.

%%%%%%%%%%%%%%%%%%%%%%%%%%%%%%%%%%%%%%%%%%%%%%%%%%%%%%%%%%%%%%%%%%%%%%%%%%

\section{Central stars}

We selected three objects from very different stellar populations,
with different metallicity and progenitor mass. The distance is known
for all three objects, allowing us to calculate the luminosity. The
results listed in Table \ref{taba} show that all three have a closely
similar value of $\log (L/L_\odot) \approx 3.7$. Because of the steep
core-mass--luminosity relation (Herwig et al. 1998), this implies that
the core masses are also very similar. The stars are hot: 70\,kK for
M\,2-31, 100\,kK for Wray\,16-423 and 140\,kK for SMP5. The
temperatures scale well with the outer radii, of 0.07\,pc, 0.09\,pc
and 0.15\,pc, respectively: the hotter the star, the more evolved the
nebula.  The ionized mass increases in the same sequence.  These
similarities suggest that the three stars are evolving on very similar
evolutionary tracks.

Gesicki \&\ Zijlstra (2000) describe a way to derive the core mass
from the dynamical nebular age and the stellar temperature. This
yields the time scale on which the central star reached its present
temperature, and this evolutionary time scale is highly sensitive to
the core mass.  The dynamical age is calculated from the mass-averaged
expansion velocity and the outer radius. A metallicity-dependent
correction is applied to allow for the slower original AGB expansion.
The resulting dynamical core masses are 0.61--0.62\,M$_\odot$ (Table
\ref{taba}), confirming that the stars are evolving on almost
identical tracks. The masses fall at the peak of the distribution for
the Bulge (Gesicki \&\ Zijlstra 2000).

The initial masses are likely different. The Bulge PN can be expected
to have an old, 1\,M$_\odot$ progenitor. The age of both Sgr PNe is
derived as 5\,Gyr (Dudziak et al. 2000), giving a progenitor mass of
$M_{\rm i}\approx 1.3$\,M$_\odot$. The value for the SMC PN is not
known, but its metallicity is very high for the SMC, which indicates a
relatively young object, with $t < 3\rm \,Gyr$ (Piatti et al. 2001,
Dolphin et al.  2001). SMP\,5 is located in the bar of the SMC, in a
region where the recent star formation (within the past Gyr) has taken
place (Van den Bergh 2000). This suggest a fairly massive progenitor.
(The sequence of initial masses is the same as the stellar $T_{\rm
eff}$, nebular radii and ionized masses.)  But the different values
for $M_{\rm i}$ have not resulted in different core masses. The core
mass at the first helium flash is almost independent of progenitor
mass, for the range $1 < M_{\rm i} <2.5 \,{\rm M}_\odot$. The
difference between this core mass and the final mass is determined by
the mass-loss efficiency (Vassiliadis \& Wood 1993, Bl\"ocker 1995).

Two of the stars show [WC]-type characteristics. However, their
stellar parameters are not obviously different from the non-[WC]
star. This lack of correlation has also been found for a study of
Galactic [WC] stars (G\'orny 2001).  Acker et al.  (2002) and Mellema
(2001) discuss the link between nebular turbulence and [WC]
hydrogen-poor winds.

The inner radii of the nebulae are very small (Table \ref{taba}). The
inner radius is determined by the pressure from the hot wind from the
central star (Mellema 1994): the small values suggest our stars have
had relatively weak winds. The present [WC] characteristics of two
objects show the presence of a stronger wind; the very hot star of
SMP5 will also have a stronger wind (Pauldrach et al. 1988). The
appearance of the nebulae suggests that these winds may have only
developed recently.  Stasinska et al.  (1998) studied PNe in five
galaxies, and concluded that, on the horizontal post-AGB/PN track, the
central stars in the Magellanic Clouds PNe evolve more rapidly than
predicted by models of Bl\"ocker (1995), which could help explain the
small inner radius.

\section{Properties of the velocity fields}

\subsection{Non-monotonic velocity structure}

The present results show evidence for complicated velocity fields
(assuming spherical symmetry: this is explored further below).
Previously, the velocity fields were modeled as monotonic increasing
with radius, approximately as $r^2$ often with strong acceleration at
the ionization front. For [WC]-type PNe, a constant velocity with
supersonic turbulence (see e.g.  Acker et al. 2002) was found. Such
simple velocity fields do not fit the PNe studied here.

It is likely that the present PNe are not unique in this.  The earlier
analyses summarized in Acker et al. (2002) were based mainly on two or
three spectral lines, and were only sensitive to simple velocity
fields. Some indications of more complicated velocity field were
noticed in Neiner et al.  (2000). They observed 8 PNe (of which 5 with
[WC] central stars) in up to 10 emission lines at a resolution of
42\,000; the data were analyzed with the Torun models.  They report
three cases where the [\ion{O}{iii}] 5007\AA\ and some other lines
are clearly not well reproduced by the models.  However, almost all
of their nebulae show smaller deviations from the models: at higher
resolution and better S/N, evidence for more complicated velocity
fields could have emerged for some of these objects as well. The two
samples together suggest that at least 50\%\ of PNe, and possibly a
much higher fraction, show evidence for more complex, non-monotonic
velocity fields.

The present sample benefits from the presence of high-quality
monochromatic images. These allow us to determine small structures in
the radial density distribution.  In the earlier analysis we very
often assumed a simplified density field because of the absence of
detailed images. For our three PNe we find indication for dense inner
shells and lower density outer regions.  The improved density
structure aids the more detailed velocity analysis.

\subsection{Turbulent velocity structure}

The presence and importance of turbulence in PNe with [WC]-type
central stars is well established. Acker et al.  (2002) fit the line
profiles of these nebulae using a constant expansion velocity, and a
turbulent component also independent of radius. It can be expected
that in reality, both components are varying with radius, but up to
now the data was insufficient to explore this further.

For one of the [WC]-type PNe in our sample, M\,2-31, we propose two
velocity solutions with the same density structure. The clearly
splitted lines of the low excitation ions exhibit an obvious Gaussian
shape, indicative of turbulence.  The wings of other lines are also
well fitted with the turbulent model.  This supports the statement of
Acker et al.  (2002) that PNe around WR planetaries suffer from
turbulent-like motions which are additional to thermal turbulence.
Nevertheless, the non-turbulent, 'wavy' solution also fits the line
profiles.  The two velocity curves are shown in Fig.\,\ref{fig_m}: the
'wavy' line follows the smooth one, and the amplitude of the
fluctuations is comparable to the turbulence value. The number of
waves is the same as the number of excitation region which we
probe. This suggests that the actual velocity structure may occur at
an even smaller scale. This situation approaches the semantics of
turbulence.  The turbulent solution is simpler and therefore more
elegant.

(The other [WC]-type PN, Wray\,16-423, shows no evidence for
turbulence. It has a weaker [WC] feature and may be classified as {\it
wels}.)

   \begin{figure}
   \resizebox{\hsize}{!}{\includegraphics{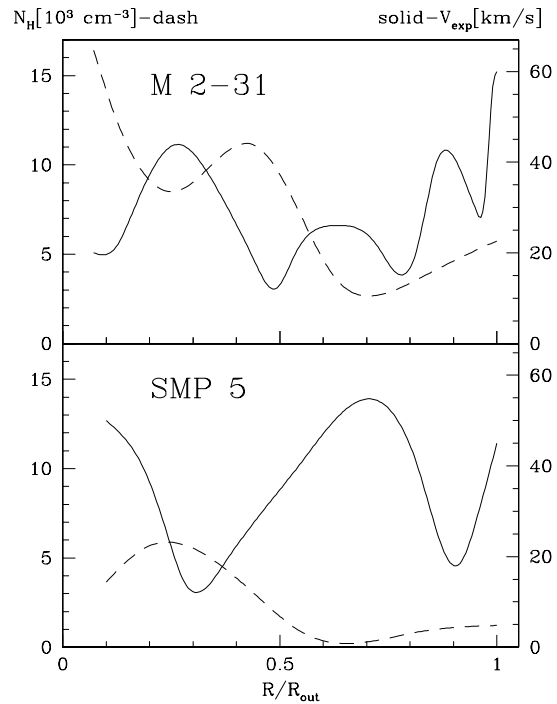}}
   \caption{The density and velocity fields for M\,2-31 and SMP\,5}
   \label{vel_ms}
   \end{figure}

The non-turbulent, 'wavy' solution shows one characteristic which the
turbulent solution by its nature hides.  Fig.\,\ref{vel_ms} shows the
density and non-turbulent velocity distribution superposed. The
velocity distribution shows an anti-correlation with the density, in
the inner regions.  A similar anti-correlation between density and
velocity is seen for the non-turbulent SMP\,5 (Fig \ref{vel_ms}).

\section{Comparison with hydrodynamics}

\subsection{Published models }

Hydrodynamical models have been developed to investigate the effects
of interacting winds in planetary nebulae (Kahn \&\ West 1985) and the
related shaping mechanisms (e.g. Mellema 1994). A three-wind
hydrodynamical model was developed by Schmidt-Voigt \&\ K\"oppen
(1987).  Sch\"onberner and collaborators have further developed these
models to study the complete evolution of PNe, including the full
mass-loss history, interacting winds, and photoionization, for
self-consistent stellar evolutionary tracks.  These are the ideal
models for comparison with the present results.  However, only one
evolutionary sequence of these highly intensive computations has so
far been published, in Perinotto et al.  (1998) (who also discuss the
history of these models) and Corradi et al.  (2000a).  The model is
calculated for a star with core mass $M_{\rm c} = 0.605\,\rm M_\odot$
(Bl\"ocker 1995), very close to the core masses of the three PNe in
this paper.  The first three models from their evolutionary sequence,
(corresponding to ages of 1776, 4603 and 5462 years counted from the
end of AGB) have sizes and densities similar to our PNe.

We briefly summarize the first phases of evolution presented in the
two mentioned papers:

\begin{itemize}

\item At epoch 1 (1776 years) the model nebula has almost
constant density ($\rho\sim 10^4$cm$^{-3}$) up to the ionization front
($r\sim10^{17}$ cm) where both density and velocity sharply increase.

\item Between epoch 1 and 2, the ionization front breaks through the
dense swept-up shell and the full circumstellar envelope ionizes.

\item At epoch 2 (4603 years) an inner denser shell
($\rho\sim5\,10^3$cm$^{-3}$) begins to develop, as a result of the
pressure from an inner 'hot bubble'. The hot bubble is due to a
shocked hot wind from the star.  The outer nebular shell,
which is the remnant of the dense shell of epoch 1, has a density
$\rho\sim2\,10^3$cm$^{-3}$; at its outer edge ($r\sim3\,10^{17}$ cm) a
shock wave is running into the ambient slow material.  The density
spike caused earlier by the ionization front, and located at the
outer edge, has almost disappeared.

\item At epoch 3 (5462 years) the central star approaches its maximum
temperature. The nebula shows a pronounced double shell structure.
The outer shell, extending to $r\sim5\,10^{17}$cm, has a nearly
constant density ($\rho\sim10^3$cm$^{-3}$): this is the swept-up AGB
wind. The inner layers have a three times higher density
with a pronounced maximum of $\rho\sim3.5\,10^3$cm$^{-3}$ 
at its inner edge ($2\,10^{17}\,$cm) and a 
much weaker maximum at its outer edge ($2.5\,10^{17}\,$cm).

\end{itemize}

\subsection{Observed  density structures }

In our PNe, the inner cavity radius is about 10\% of the outer shell
radius.  In the hydrodynamical models the inner radius of ionized
nebula is about 50\% of the outer radius.  This difference suggests
that the transition from AGB to post-AGB wind occurred different from
Sch\"onberner's scenario.  In subsection 7.1 we suggested that the
fast wind may have been weaker than assumed in the hydrodynamical
models. Alternatively, the star may have evolved faster (the core mass
is slightly higher than in the models) giving the fast wind less time
to act, or the AGB expansion velocity may have been low giving a
denser, more compact AGB shell. Perinotto et al. (1998) discuss two
sequences differing only in the initial matter distribution. Two
snapshots from these sequences, taken at about 1800 years, show
a markedly different density distribution. Our experience suggests that
the density and velocity distributions would be discernible with the
Torun models.

The structure of our PNe, with a high density inner ring and a low
density outer region, corresponds to theoretical models 4--5\,$10^3$
years after the AGB.  But the models are density bounded while our PNe are
ionization bounded. Our PNe have inner rings denser and closer to the
central star than the theoretical models.  It is possible that the
growing density of the inner shell blocked the ionizing radiation and
a previously fully ionized PN entered a recombination phase.

Apart from this difference, the epoch-2 model shows a structured dense
inner ring with an outer region in which the density slightly
increases outwards.  Such a situation we met in SMP\,5.  In the
epoch-3 model there is a main density maximum at the inner nebular edge
with a smaller but noticeable density bump further out.  This
situation we found in M\,2-31.

\subsection{Velocity structures: Making waves}

For the first time our analysis led to strongly-non monotonic
velocities. It is encouraging that the velocity spikes which we find
correspond to features found in hydrodynamical models. However, the
intermediate velocity jump can also be interpreted in terms of clumpy
structures or asphericity, as discussed later.

Two PNe show a velocity
field in a 'W' shape.  The velocity distribution for SMP\,5 is
anti-correlated with the density run (see Fig.\,\ref{vel_ms}) as can
be expected for gas flows.  
The hydrodynamical models suggest several processes are responsible for
the features in the velocity field.

\begin{itemize}

\item The innermost region represents the contact discontinuity
between the hot inner region and the PN. It is accelerated by the fast
wind of the central star. For a typical mass loss rate of $\dot M =
10^{-9} \, \rm M_\odot\,yr^{-1}$ and velocity $v = 10^4\,\rm
km\,s^{-1}$, the energy in the accelerated shell in our models
equates to roughly 400\,yr of the wind energy. The acceleration can
occur through thermal pressure (Sch\"onberner \&\ Steffen 1999) or
through direct acceleration of a wind-swept shell (Harrington et
al. 1995).

\item The outer velocity increase is well understood, and represents 
the shock driven into the ambient AGB wind by the thermal pressure of
the ionized planetary nebula.

\item The intermediate shock seen as a weak feature in Sch\"onberner
\&\ Steffen (1999) is the compressed AGB wind (a density enhancement)
at the outer edge of the inner dense swept-up shell.

\end{itemize}

The outer velocity increase is seen in all three objects. The innermost
increase is seen in two objects; in the third object (M\,2-31) it
falls in the \ion{He}{iii} region for which we have no velocity probe. 
The explanation for these apparently common features appears secure.

The intermediate shock is seen in the same 2 objects (and in M\,2-31
may take the form of a superposition of many such shocks). It is
not clear that the above explanation as the compressed AGB wind
is unique. 

Mass-loss variations in the AGB wind could cause density jumps at
various radii. An effect of such a jump is in fact present in the
models of Corradi et al. (2000a), namely the trace of the last thermal
pulse on the AGB, which in the model sequence is located a distance of
about $10^{18}$\,cm.  The velocity jump associated with this feature
grows to about 20\,km\,s$^{-1}$ as the main expanding nebula
approaches this structure. This particular feature is not observable,
being located in the low-density AGB wind. However, its evolution
suggests that any mass-loss fluctuation during the ejection of the
main nebula may affect the density and velocity field in the
PN. Evidence is emerging that the AGB mass loss fluctuates on time
scales of the order $10^2$--$10^3$\,yr (Hashimoto 1994, Zijlstra et
al.  2002).

The turbulence in M\,2-31 could be interpreted as a special case where
instead of a single intermediate shock, many shock-lets are
present. The winds from [WC] stars show strong time-variability
(e.g. Grosdidier et al. 2000): the variable pressure at the inner edge
of the PN could cause a progression of shock waves traveling into the
nebula.  The velocity amplitude of the observed waves, or the
turbulent velocity, is little larger than the sound speed ($\sim 12
\rm \, km\,s^{-1}$). This suggests an unusual picture of a planetary
nebula criss-crossed by a multitude of sound waves.

\section{Alternative explanations: elliptical and clumpy nebulae}

We should consider the possibility that in our PNe the inner shock
found can actually be a spherical approximation to more azimuthally
irregular situation. The nebulae can be elliptical in shape exhibiting
different structure in different directions. The wind breaking through
the holes in the inner dense shell can accelerate a layer at
intermediate radii, and the inner nebular region may be a mixture of
faster and slower matter. The outer shell can fragment into clumps.

A 3D photoionization modelling which can 
address such situations has been applied to NGC\,3132
by Monteiro et al. (2000). Such modelling usually attempts to fit the
observed image of a nebula. Monteiro et al.  reached an important
conclusion that different geometries can reproduce the same
observational data. The low-resolution spectroscopic data can be well
modelled by an ellipsoidal geometry, but by using additional data from
[\ion{S}{ii}] line ratios and [\ion{O}{iii}] velocity profiles they
deduce a Diabolo-shape geometry for NGC\,3132. However, in their
calculations they made several simplifications, among them the
assumption of a linearly increasing velocity field. Such modelling
appears promising but at present still suffers from the large number
of free parameters involved, and from the prohibitive CPU time needed
for fully self-consistent calculations; because of these limitations,  
spherical models still should
be considered as valuable approximations.

We discuss briefly two possibilities which may affect the discussion
of Wray\,16-423 and SMP\,5 (the turbulent M\,2-31 is a different
story). 

\subsection{Elliptical nebulae}

   \begin{figure}
   \resizebox{\hsize}{!}{\includegraphics{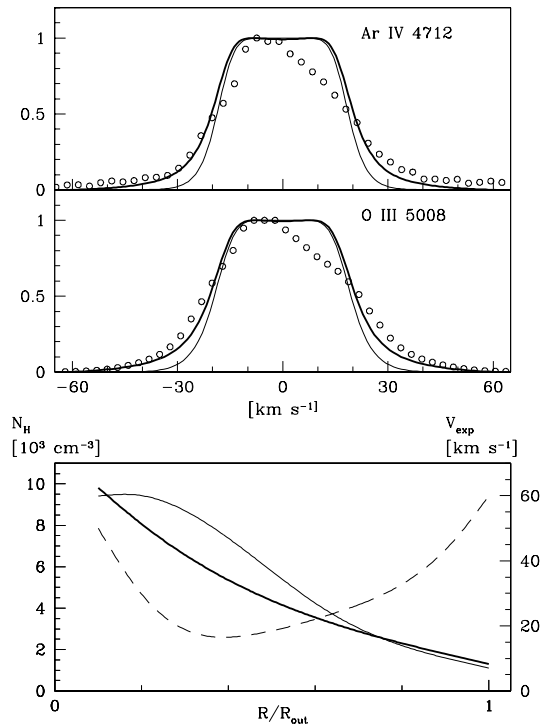}}
   \caption{Two different density fields for Wray\,16-423 resulting in
different line shapes. Two upper panels present the emission line
shapes for [\ion{O}{iii}] and [\ion{Ar}{iv}] ions obtained with the
same velocity field (dashed line in lower box) but with two density
distributions (thin and thick lines in lower box).}
   \label{elipt}
   \end{figure}

We first investigate whether the central velocity cusp of the 'W'
could be an artifact of the assumption of spherical symmetry.  This is
done for the case of Wray\,16-423 where some evidence for ellipticity
exists.  The requirement for a high velocity region  comes from
the extended wings of the [\ion{O}{iii}] and [\ion{Ar}{iv}] lines:
we therefore attempt to reproduce these without such a region.

To approximate an elliptical model, we use a two-component model.
This essentially doubles the number of free parameters, so that one
can expect that the data can be fitted much more easily but the
uniqueness of the model will be affected. To limit the number of free
parameters, we fix the inner and outer radius, and the velocity
profile, and only vary the density distribution.  Compared to our
spherical model, one of the components must be opaque with an
ionization boundary at our estimated outer radius. This component will
explain the observed low excitation lines. The second component has a
lower density, giving a transparent model with the same outer
radius. (A different outer radius could be assumed but this would be
wholly unconstrained with the present data.)

 We adopt a 'U'-shaped velocity field which is the same in all
directions, as found observationally for NGC\,7027 (Walsh et
al. 1997a).  Our question is now whether a density profile exists for
the transparent component which can reproduce the line wings. One such
possibility is presented in Fig.\,\ref{elipt}.  This shows that a
relatively small change in density distribution will shift the line
formation zones of [\ion{O}{iii}] and [\ion{Ar}{iv}] ions outwards so
that the line wings trace the outermost acceleration.  The density
should in fact not be too much lower, otherwise the line wings
would be too weak to be observed.  Not shown are the
\ion{He}{ii} line which is also well fitted and other observed low
excited lines which are not fitted. 

This example suggests a possible explanation of our observations, not
in terms of a variable velocity in the intermediate region, but a
variable azimuthal ionization stratification with a simpler velocity
field. This discussion certainly concerns also SMP\,5. The turbulent
velocity field of the third PN could not easily be fitted in this way.

It is well known (Gruenwald et al. 1997) that without spherical
symmetry the outward-only approximation fails. This approximation is
applied in our photoionization code. Therefore we refrain from
quantitative analysis. The tentative model is only intended to give
support for the considered possibility.

\subsection{Clumps and irregularities}

PNe show a wealth of sub and microstructures, including clumping. The
best known case is perhaps the Helix, but HST shows small structures
in many objects. However they are far easier seen in  images than
in spectra. High density clumps tend to have lower
temperatures (e.g. the dusty Helix globules observed by Meaburn et
al. (1992), low ionization knots seen in images of Balick (1987) and
discussed by Kahn \& Breitschwerdt 1990), while the forbidden lines
are highly temperature sensitive and tend to trace the extended gas.

Large and bright knots are also observed in some nebulae. However, the
case of NGC\,6543 analyzed by Bryce et al. (1992) shows that the
explosive appearance seen in images in fact  corresponds to a relatively
inert kinematic situation where the bright knots are co-moving with the
faint halo. Also Corradi et al. (2000b) when analyzing nebulae IC\,2553
and NGC\,5882 found that knots visible in low-ionization lines do not
show evidence of moving with peculiar velocities compared with the
general motion.

However,  clumps can affect the velocity structure.  If the inner dense
shell is clumpy, the fast wind from the central star could be breaking
through the holes, and set up a secondary swept-up shell at a larger
radius, in the lower density outer regions.  Such an effect is seen in
the Abell nebulae A30 and A78, which have dense clumps very close to
the star, with a cometary appearance, and an irregular, accelerated
shell at about 0.5 of the outer radius (Meaburn et al.  1998, Meaburn
\& L\'opez 1996). In these objects the location of the fast wind is
obvious in observations, as it is hydrogen-poor.

It is also possible that the outer shell is clumpy, e.g., as in the
already mentioned IC\,2553 and NGC\,5882. Both PNe reveal a double
shell structure; both outer shells are more spherical and expand
faster than the inner shells. The low ionization knots, of intriguing
origin, follow the movement of the surrounding outer shells. This also
gives a situation, as in the elliptical model above, where the same
accelerated outer regions can explain both the extended wings of high
excited lines in the transparent fragments, and the line splitting of
the low excited lines in the opaque fragments.

\section{Conclusions}

High-resolution echelle observations, covering a range of emission
lines, provide a powerful tool to study the velocity distribution
inside (unresolved) planetary nebulae. We analyze three objects, in
the Bulge, the Sagittarius Dwarf galaxy and the SMC.  They show
similar structure: a small inner radius, a dense inner region and a
lower-density outer shell. The dynamical core masses are almost equal,
around 0.62\,M$\odot$, in spite of very different metallicities and
progenitor masses.  All three PNe show evidence for jumps in expansion
velocity, at various locations in the nebulae.  The velocity
stratification as found in our spherical model is supported by the
observed ion stratification.  In Wray\,16-423 and M\,2-31 we detect
different line splitting for different ions, confirming that the low
excitation ions trace more extended regions.

The proximity of the dense inner shell to the central star is not
seen in  the existing hydrodynamical models. The assumptions
concerning late AGB and early post-AGB mass-loss evolution need to be
verified to model the three presented PNe.

Three main velocity features appear. Velocity maxima are found at the
inner and the outer edge of the nebulae. The outer acceleration is a
common, known feature; the evidence for the innermost acceleration is
new. (In one case, M\,2-31, the evidence at the inner edge is
tentative as we lack a good probe for the region.)  Hydrodynamic
models, and energy arguments, strongly suggest that the inner one is
due to acceleration by the fast wind from the hot central star. The
outer one can be explained as the overpressured ionized region pushing
into the outer AGB wind. The third feature is a high velocity, low
density region affecting the intermediate excitations ions, which is
seen in two objects. The origin of this region is not clear. Three
possible explanations are given, all of which have some support in
models and/or observations of PNe.  First, it can be associated with a
weak shock seen in hydrodynamical models at the boundary where the
dense inner shell expands into the lower-density region. However, the
observed feature is much stronger than in the model.  Second, it can
be due to an ellipticity of the nebula, where in a direction with
lower density, the line formation region of the intermediate excitation
ions extends into the outer
acceleration zone. A similar situation could exists if the outer
region is clumpy. Thirdly, the region could trace a secondary shell
accelerated by a stellar wind breaking through an innermost clumpy
region, as seen in, e.g., the PNe A30 and A78.  It is not possible to
decide on the correct explanation based on the present data.

One nebula shows evidence for either fine radial velocity structure,
or  strong turbulence. This PN has a [WC]-type central
star, a characteristic which has previously been linked to turbulent
nebulae. There are indications for an anti-correlation between density
and velocity. We suggest that the apparent turbulence may be a
superposition of many shocks with velocity near the sound speed.  Such
waves could perhaps be caused by a time-variable [WC]-wind.

\section*{Acknowledgments}
K.\,Gesicki acknowledges support from 'Polish State Committee for
Scientific Research' grant No. 2.P03D.020.17. The observer
at the NTT was David Pinfield. The research was supported by grants from
the British Council and by a PPARC Visitors Grant.

\end{document}